\documentclass[aps,prl,reprint,floatfix,longbibliography,twocolumn]{revtex4-2}
\usepackage{amsmath,amssymb,amsthm,hyperref,graphicx}
\hypersetup{hidelinks}
\usepackage[capitalise]{cleveref}
\hypersetup{colorlinks=true, linkcolor=blue, citecolor=red, urlcolor=blue}
\begin{document}
\title{Using surface plasmons to detect spin inertia}
\author{H. Y. Yuan}
\email{Contact author: hyyuan@zju.edu.cn}
\affiliation{Institute for Advanced Study in Physics, Zhejiang University, 310027 Hangzhou, China}

\date{\today}

\begin{abstract}
Recent experiments demonstrate that spin dynamics may acquire an inertial effect in a few metallic magnets, deviating from the traditional inertia-free dynamics. It remains an open question to ascertain the physical mechanisms and universality of the spin inertia across diverse magnetic systems. Here, we show that spin inertia generates nutation spin waves in the terahertz regime, which can hybridize with the surface plasmons in two-dimensional (2D) conducting materials such as graphene. By exciting hybrid spin wave-plasmon modes and analyzing the reflection spectrum of a 2D material$|$magnet heterostructure, we propose a method to quantitatively determine the strength of spin inertia in magnetic layers. Our approach is universally applicable to all types of magnetic insulators and could advance the future exploration of the magnitude and physical mechanism of spin inertia.
\end{abstract}

\maketitle
\textit{Introduction.--} Understanding spin dynamics is central to magnetism and spintronics. Since the pioneering work of Landau, Lifshitz, and Gilbert (LLG), it has been gradually accepted that spin dynamics can be well described by the phenomenological equation \cite{Landau1935,Gilbert1956}
\begin{equation}\label{llg}
\partial_t \mathbf{m} = - \gamma \mu_0 \mathbf{m}\times \mathbf{H}_\mathrm{eff} + \alpha \mathbf{m}  \times \partial_t \mathbf{m},
\end{equation}
where $\mathbf{m} =\mathbf{M}/M_s$ is the normalized magnetization vector with $M_s$ being saturation magnetization, $\mu_0$ is vacuum permeability, $\gamma$ is gyromagnetic ratio, and $\mathbf{H}_\mathrm{eff}$ is the effective field which may include exchange field, anisotropy field, dipolar field, and external field applied to a magnet. The first and second terms on the right-hand side of Eq. \eqref{llg} represent the precessional and damping motion of magnetization around the effective field, respectively. The Gilbert damping $\alpha$ characterizes the dissipation strength. The first-order time derivative of magnetization $\partial_t \mathbf{m}$ in the LLG equation implies inertia-free dynamics, distinguishing it from classical spinning tops with inertial effects such as nutation \cite{Wegrowe2012}.

\begin{figure}
	\centering
	\includegraphics[width=0.48\textwidth]{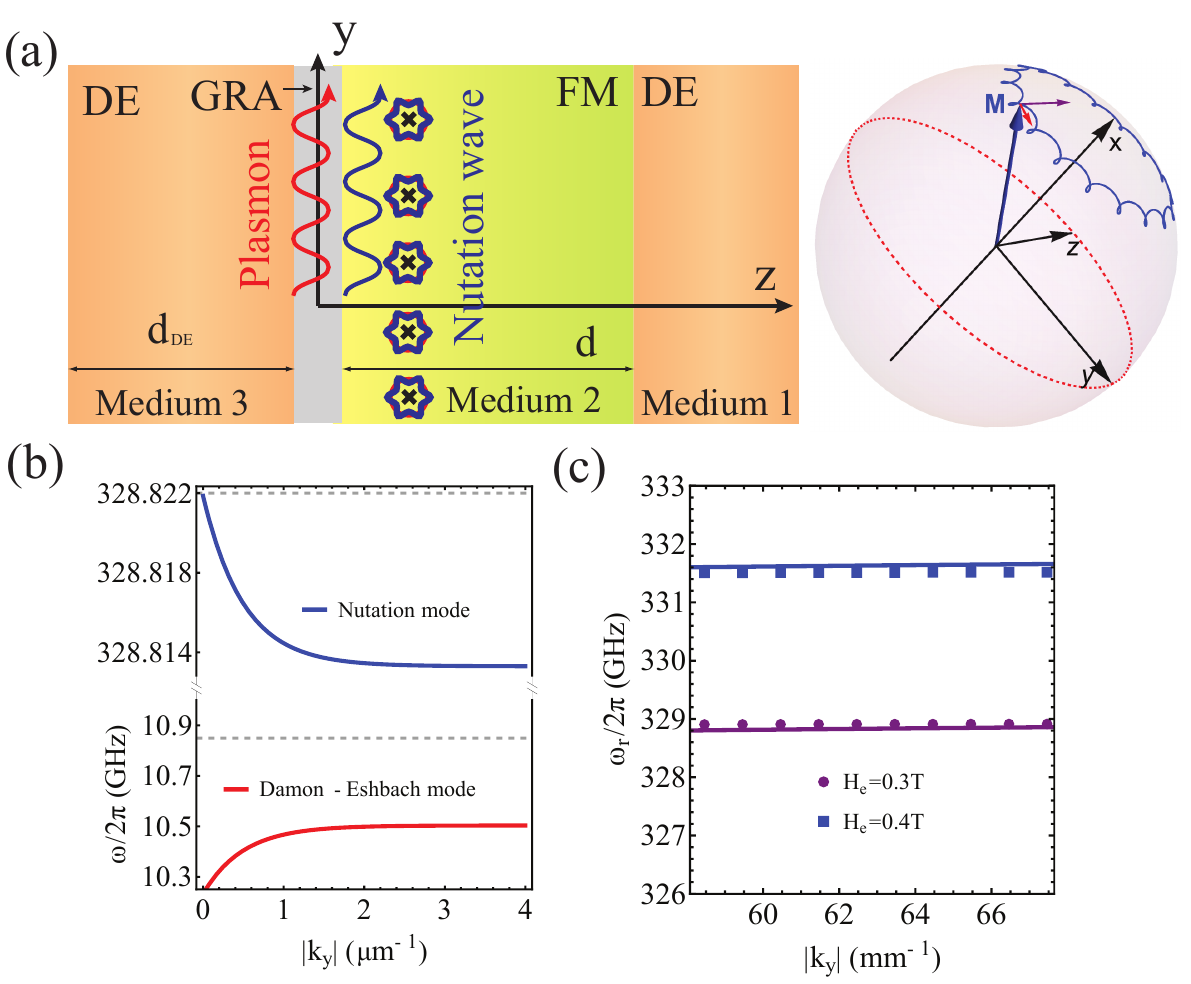}\\
	\caption{(a) Cross-sectional view of a DE$|$GRA$|$FM$|$DE hybrid structure with $z$-axis being the thickness direction. The nutation spin waves and surface plasmons are coupled at the interface. The right panel sketches the nutation motion (blue curve) superimposed above the normal precession of magnetization on a sphere with magnetization length as the radius. The red and purple arrows sketch the directions of precessional and nutational torques, respectively. (b) Dispersion relation of the Damon-Eshbach mode (red line) and nutation surface mode (blue line). Parameters of yttrium iron garnet (YIG) are used: $\mu_0M_s=0.175~\mathrm{T},~\mu_0H_e=0.3~\mathrm{T},~d=1~\mu m$, $\eta=0.5~\mathrm{ps}$. (c) Dispersion relation of the hybrid plasmon-spin wave modes. $E_F=0.1~\mathrm{eV}$. The symbols are numerical solutions of Eq. \eqref{dispersion_hybrid_infite_fm} while the lines represent the analytical results calculated based on Eq. \eqref{analytical_hybrid_excitation}.}\label{fig1}
\end{figure}

While the LLG framework including various spin torques successfully explains experimental results of magnetic resonance \cite{KittelPR1948}, magnetization switching \cite{Slon1996,Berger1996,FukamiNN2016}, spin wave excitation \cite{ChumakNP2015,YuanQM,YuPR2021} and magnetic textures dynamics \cite{Mougin2007}, recent experiments reveal anomalous subterahertz (THz) magnetization response to narrow band THz fields \cite{Neeraj2021,Unikand2022,YLi2015}, which is much higher than the typical ferromagnetic resonance frequency in the gigahertz (GHz) regime. To account for this, an additional inertial term $\eta \mathbf{m} \times \partial_{tt} \mathbf{m}$ is introduced to the original LLG equation, where $\eta$ quantifies the nutation timescale. The new inertial term generates nutation motion of spins superimposed on the precessional trajectory (right panel of Fig. \ref{fig1}(a)), and the value of $\eta$ characterizes the typical timescale of nutation oscillations. The expected values of the nutation time $\eta$ should be in the subpicosecond scale \cite{Neeraj2021, Unikand2022} in order to explain the high-frequency response of magnetization in various materials. The ultrafast timescale of spin inertia holds potential applications for novel data processing technologies and thus attracted significant attention.

Notably, the inertial term $\mathbf{m} \times \partial_{tt} \mathbf{m}$ preserves the time-reversal symmetry and does not contribute to energy relaxation of the magnetic system, unlike the Gilbert dissipation term. Multiple theories have been proposed to understand the underlying mechanisms of spin inertia. However, experiments and theories predominately focus on metallic magnets \cite{Neeraj2021,Ciornei2011,Unikand2022,YLi2015,Fähnle2011,Bhat2012,Danny2015,Thonig2017,Kikuchi2015,Mondal2017,Mondal2018,Mikhail2021,Mondal2023,HePRB2024}, where the interaction between electrons and spins plays an important role to induce the inertial effects. On a microscopic level, F\"{a}hnle et al. showed that spin inertia may naturally arise within the framework of extended breathing Fermi surface model \cite{Fähnle2011}. Bhattacharjee et al. derived a general expression of spin inertia in the spirit of an $s$-$d$-like interaction between the electron spins and local magnetization \cite{Bhat2012}, while the nonadiabatic effect from the environmental degrees of freedom was also discussed \cite{Kikuchi2015}. Thonig et al. provided a theoretical model for calculating the spin inertia within torque-torque correlation model based on \textit{ab-initio} electronic structure footings \cite{Thonig2017}. Mondal et al., attributed the spin inertia to higher order relativistic effects from a fundamental Dirac Hamiltonian \cite{Mondal2017, Mondal2018}.

Whether the inertial spin dynamics is a universal effect for magnetic insulators still lacks experimental evidence, despite some theoretical predictions that does not relying on conducting properties of the systems \cite{Ciornei2011,Wegrowe2012,Titov2021,Andres2022,Mario2024}. Ciornei et al. derived the generalized LLG equation with spin inertia based on the mesoscopic nonequilibrium thermodynamics theory \cite{Ciornei2011,Wegrowe2012}. Quarenta et al. showed that low and high frequency modes in the environment of spins can universally recover the Gilbert damping and inertial terms in the LLG equation \cite{Mario2024}, where the environment may include electrons, phonons or other bath modes. Addressing the universality of spin inertia question is of particular interesting and importance because low-damping magnetic insulators are promising in spintronic applications. Mediating the gap between metals and insulators requires more flexible method to excite and detect the spin inertial effects, which motivates our current work.

In this work, we show that nutation spin waves can drag the excitation of surface plasmons in a graphene$|$magnet hybrid structure. The THz nature of nutation spin waves perfectly align with the surface plasmon frequency in graphene. Thanks to the high conductivity of graphene, the excitation of surface plasmons carries away a significant amount of electromagnetic energy and induces a dip in the reflection spectrum. By calibrating the dip position, one can precisely determine the strength of nutation in the magnetic materials. Our proposal applies broadly for a wide range of magnetic insulators and should benefit the future study of nutation effects in magnetic systems. The combination of plasmons and nutation spin waves may offer novel opportunities for nanophotonics and ultrafast spintronics.

\textit{Hybrid nutation spin wave-plasmon excitation.}---We consider a graphene$|$insulating ferromagnet (GRA$|$FM) heterostructure sandwiched between two dielectric (DE) layers as shown in Fig. \ref{fig1}(a). The theoretical framework to study the hybrid spin wave-plasmon excitation was developed in Ref. \cite{YuanPRL2024,YuanPRB2025}. In general, the electromagnetic wave inside a magnet should satisfy the Maxwell equation 
\begin{equation}\label{MaxwellHM}
(\nabla^2 +k^2) \mathbf{H}- \nabla(\nabla \cdot \mathbf{H}) + k^2\mathbf{M} =0,
\end{equation}
where $k^2 = \epsilon \mu_0 \omega^2$ is the wavevector of the electromagnetic wave with $\epsilon$ and $\omega$ being the relative permittivity and angular frequency of the wave, respectively. $\mathbf{M}=M_s\mathbf{m}$ is the magnetization of the magnetic layer.
The magnetization relates to the magnetic field $\mathbf{H}$ as $\mathbf{M}=\chi \cdot \mathbf{H}$. To incorporate the inertial effect, we derive the magnetic susceptibility $\chi$ from the inertial LLG equation 
\begin{equation}\label{inertial_LLG}
\partial_t \mathbf{m} = - \gamma \mu_0 \mathbf{m}\times \mathbf{H}_\mathrm{eff} + \mathbf{m}  \times (\alpha \partial_t \mathbf{m} + \eta \partial_{tt}\mathbf{m}). 
\end{equation} 
Here the effective magnetic field $\mathbf{H}_\mathrm{eff}$ includes the exchange field ($2A\nabla^2\mathbf{m}$) with $A$ being the exchange stiffness, external magnetic field $H_ee_x$ and dipolar field $\mathbf{H}$, i.e. $\mathbf{H}_\mathrm{eff}=2A\nabla^2\mathbf{m} + H_ee_x +\mathbf{H}$.
In the calculation, we consider the spin wave excitation above the ground state $\mathbf{M}_0=M_se_x$ and expand the magnetization as $\mathbf{M}=\mathbf{M}_0 + M_ye_y + M_ze_z$ with $M_{y,z} \ll M_s$ and the dynamics of $(M_y,~M_z)$ is derived by linearizing the inertial LLG equation \eqref{inertial_LLG} as
\begin{equation}\label{llgmymz}
\left (\begin{array}{c}
  M_y \\
  M_z
\end{array} \right )=
\left ( \begin{array}{cc}
  \kappa & -i\nu\\
  i \nu & \kappa
\end{array} \right )
\left (\begin{array}{c}
  H_y \\
  H_z
\end{array} \right ),
\end{equation}
where 
\begin{equation} \label{suskv}
\kappa=\frac{\omega_k \omega_m}{\omega_k^2 - \omega^2},~~\nu=\frac{\omega_m \omega}{\omega_k^2 - \omega^2}.
\end{equation}
Here, $\omega_k=2Ak^2 + \omega_h-i\alpha \omega - \eta \omega^2$ with $\omega_h=\gamma \mu_0 H_e$, $\omega_m = \gamma \mu_0 M_s$ and $A$ being the exchange stiffness. Note that the inertia term $\eta$ enters our magnetic susceptibility and will influence the magnetic response to incident electromagnetic wave as we shall see below. Here the magnetic damping $\alpha$ will give a modification to the resonance frequency on the order of $\alpha^2$ \cite{GMbook}, which is usually very small for magnetic insulators.

Now we have a complete set of equations to describe the electric and magnetic excitations in the hybrid structure, which allows us to find the electromagnetic waves in both the dielectric and magnetic regions. The surface plasmons in the 2D layer generate a surface electric current $\mathbf{j}=\sigma \mathbf{E}_3(z=0)$ connecting the magnetic fields at the interface of media 3-2 \cite{Peresbook}, where $\sigma$ is the alternating conductivity of the 2D material. In the THz regime, $\sigma = 4\sigma_0 E_F/(\pi\Gamma - i\pi\hbar \omega)$ with $\sigma_0= e^2/4\hbar$, $E_F$ being the Fermi energy and $\Gamma$ being the relaxation rate of carriers. Here the prefactor 4 comes from the double degeneracy of spin and valley in graphene \cite{Mikhailov2007}. By matching the continuity of tangential electric and magnetic fields on the two surfaces of the magnet ($z=0,d$),  we derive the dispersion equation of the hybrid mode as (See Appendix for the details of derivation)
\begin{equation}\label{dispersion_hybrid}
\begin{aligned}
\left ( \frac{k_2^4}{\kappa_1 \kappa_3} - \delta^-\delta^+ \right ) \sinh(k_yd) + \frac{k_2^2}{2\kappa_3} (e^{-k_yd}\delta^+-e^{k_yd}\delta^-)\\
\frac{k_2^2(1-\sigma')}{2\kappa_1}(e^{-k_yd}\delta^--e^{k_yd}\delta^+)+\sinh(k_yd)\delta^-\delta^+ \sigma'=0
\end{aligned}
\end{equation}
where $k_i^2 = \omega^2 \epsilon_i/c^2$, $\kappa_i = \sqrt{k_y^2-k_i^2}$ is the decay wavevector with $1/\kappa_i$ being the skin depth of electromagnetic wave, $k_y$ is the prorogating wavevector of electromagnetic wave, $\sigma'=i\mu_0\omega \sigma/\kappa_3$, $d$ is the thickness of the magnetic layer, and $\delta^\pm$ represents the response of the magnet to the incident electromagnetic wave
\begin{equation}\label{deltapm}
\delta^\pm = \pm \frac{k_2^2}{k_y}\frac{1+\kappa \pm \nu}{1\mp \nu k_2^2/k_y^2}.
\end{equation}
Here we emphasize that Eq. \eqref{dispersion_hybrid} fully includes the dissipation of magnetic spins by the Gilbert damping in the magnetic susceptibility elements $\kappa$ and $\nu$ and the relaxation of surface of plasmons in the real part of graphene conductivity $\sigma$.

Let us first consider a single ferromagnet placed in the air, i.e. $\sigma'=0,~\epsilon_1=\epsilon_3=1$, the dispersion is reduced to
\begin{equation}
\begin{aligned}
\left(1+\nu \frac{k_2^2}{k_y^2}\right)\left(1-\nu \frac{k_2^2}{k_y^2}\right) + (1+\kappa + \nu)(1+\kappa - \nu) \\
-2(1+\kappa + \nu \frac{k_2^2}{k_y^2})\coth(k_y d)=0.
\end{aligned}
\end{equation}
This is the precise result of spin wave dispersion taking into account of both electric and magnetic components of the electromagnetic wave. 
In the magnetostatic limit, $k_2^2/k_y^2 = \omega^2/(c^2k_y^2) \ll 1$, one formally recover the literature result in the absence of nutation \cite{DE1960, DE1961,Stancil2021}
\begin{equation}
2(1+\kappa)(1-\coth(k_y d)) + \kappa^2 -\nu^2=0.
\end{equation}

By incorporating the magnetic susceptibility \eqref{suskv}, we can derive the spin wave dispersion in the absence of magnetic damping ($\alpha=0$) as
\begin{equation}
\begin{aligned}
\eta^2 \omega^4 - (2\eta \omega_h + \eta \omega_m + 1) \omega^2 \\
+\omega_h^2 + \omega_h \omega_m + \frac{\omega^2}{2(1-\coth (k_y d))}=0.
\end{aligned}
\end{equation}
Here we neglect the exchange coefficient ($A=0$) because it only makes a tiny correction to the susceptibility at considered wavelength. The influence of the exchange interaction on the spectrum appears at the short wavelengths as a quadratic dependence of eigenfrequencies. This effect will be studied elsewhere. Note that the nutation motion is very fast compared to the typical precession frequency, i.e. $\eta \omega_{m,h}\ll 1$, the two solutions to this equation can be explicitly written as
\begin{subequations}\label{dispersion_nsw}
\begin{align}
\omega_+ &\approx \frac{1}{\eta} \left[  1 + \left (\omega_h + \frac{\omega_m}{2} \right ) \eta + \frac{1}{8} \omega_m^2 \eta^2 e^{2k_yd} \right ],\\
\omega_- &\approx \sqrt{\left (\omega_h + \frac{\omega_m}{2} \right)^2 - \frac{1}{4} \omega_m^2 e^{2k_yd}}.
\end{align}
\end{subequations}

It is clear that $\omega_-$ is the normal Damon-Eshbach (DE) mode \cite{DE1960,DE1961,Serga2010} while $\omega_+$ is the surface spin wave mode contributed by the nutation effect. Note that the sign of $k_y$ depends on the direction of magnetization ($e_m$) and surface normal ($e_n$) according to the relation $\mathbf{e}_m \times \mathbf{e}_k = \mathbf{e}_n$ \cite{DE1960,WangPRL2020}. In our setup shown in Fig. \ref{fig1}(a), the surface wave propagates toward the negative (positive) $y-$axis at the surface $z=0$ ($z=d$), respectively. Figure \ref{fig1}(b) shows the typical spectrum of Damon-Eshbach mode and nutation surface spin wave mode. Here we notice that the nutation effect will slightly suppress the frequency of DE mode at very large $|k_y|$ (dashed line in Fig. \ref{fig1}(b)). This is because the nutation term will generate nutation motion above the normal precession motion, elongating the moving orbit and precession time, as shown in the right panel of Fig. \ref{fig1}(a). In the long-wavelength limit, i.e. $k_y=0$, we recover the frequency of uniform nutation resonance, i.e. $\omega^+ \approx 1/\eta + \omega_h +\omega_m/2$, the additional term $\omega_m/2$ comes from the contribution of dipolar interaction, which is not included in the literature  \cite{MikhailPRB2020, MondalPRB2021,Mondal2023}.


Now let us consider the hybrid magnon-plasmon excitation. In general, the dispersion equation \eqref{dispersion_hybrid} is difficult to be solved analytically and one has to rely on numerical results. Here, to get some intuition of the hybrid modes, we consider a thick ferromagnet, such that $|k_y d|\gg 1$ and the $e^{|k_y|d}$ terms dominate Eq. \eqref{dispersion_hybrid}. Moreover, the electromagnetic boundary conditions on the interface of media 2-1 disappear, implying the absence of $\kappa_1$ terms in Eq. \eqref{dispersion_hybrid}. Then we come to the simplified dispersion equation
\begin{equation}\label{dispersion_hybrid_infite_fm}
k_2^2/\delta^- + \kappa_3 -i \mu_0 \omega \sigma =0. 
\end{equation}

Now we focus on the THz regime, which matches the typical nutation frequency. Since the relaxation time of spin waves is usually long enough to observe the effects of wave propagation, we may first neglect the damping term in $\kappa,~\nu,~\delta^-$. \textcolor{red}{} Considering the magnetostatic limit, i.e. $\kappa_3 =\sqrt{k_y^2-\omega^2 \epsilon_3/c^2} \approx |k_y|$ and the neglecting the small magnetic damping, Eq. \eqref{dispersion_hybrid_infite_fm} is reduced to a simple quadratic equation
\begin{equation}
\eta \omega^2 - \omega - \omega_h - \frac{\omega_m (4\mu_0E_F\sigma_0/(\pi \hbar)-k_y)}{4\mu_0E_F\sigma_0/(\pi \hbar)-2k_y}=0.
\end{equation}
We can analytically solve the equation and derive the dispersion relation of the hybrid excitation as
\begin{equation}\label{analytical_hybrid_excitation}
\omega_\mathrm{mp} = \frac{1}{\eta} + \omega_h+ \frac{\omega_m }{2} + \frac{\mu_0E_F\omega_m\sigma_0}{2\mu_0E_F\sigma_0 - \pi\hbar k_y}.
\end{equation}
Here the first term $1/\eta$ on the right-hand side makes a leading contribution to the frequency, which is approximately equal to the frequency of the nutational spin wave mode in Eq. \eqref{dispersion_nsw}.
Figure \ref{fig1}(c) shows the dispersion relation of hybrid plasmon and magnon excitations, our analytical result can well capture the main feature of the numerical result by fully solving \eqref{dispersion_hybrid_infite_fm}. We further emphasize that a rigorous mathematical treatment of the imaginary part of the frequency when damping of plasmons and magnons are included would be beneficial to understanding the fine dynamics of the system \cite{Kukh2015,Silva2015,Kukh2024,YuanPRB2025}, however, the real spectrum of hybrid magnon-plasmon mode is still robust against the influence of dampings. The hybrid excitation can be clearly identified in the single-valley structure in the reflection structure of the system despite the influence of dampings as we shall see below, which should make it feasible to be observed in experiments.

\begin{figure}
	\centering
	\includegraphics[width=0.48\textwidth]{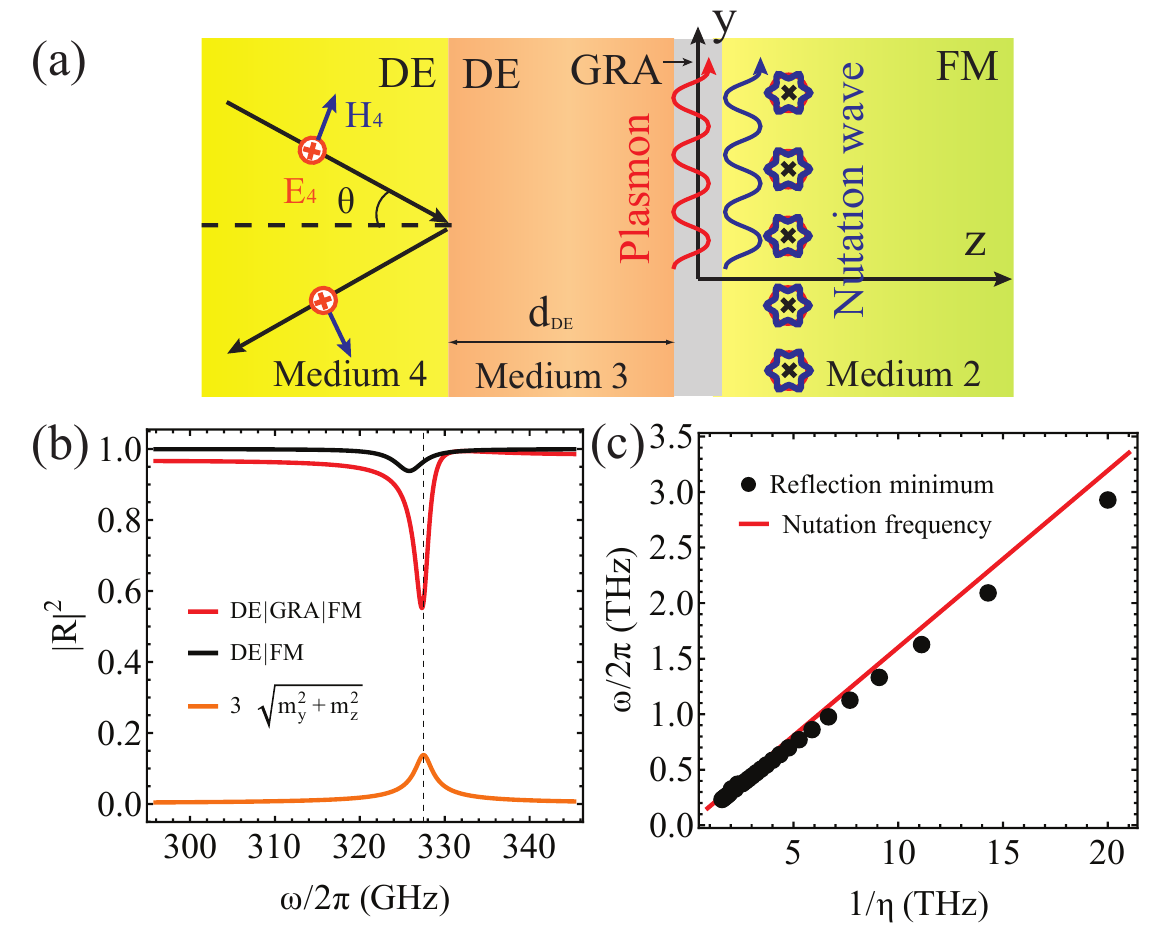}\\
	\caption{(a) Scheme of a DE$|$DE$|$GRA$|$FM hybrid structure. An electromagnetic wave inputs from medium 4 and penetrates into medium 3 as an evanescent wave, exciting the surface plasmons and spin waves at the interface of media 3 and 2. (b) Reflection spectrum of the system as a function of incident wave frequency. The orange line plots the amplitude of spin waves.  (c) Local minimum position of the reflection spectrum as a function of nutation time $\eta$. The red line plots the theoretical nutation frequency according to Eq. (11a). $\theta=-67.58^\circ,~\eta=0.5~\mathrm{ps},d_{DE}=5~\mu m, \epsilon_4=14,\epsilon_3=2,\epsilon_2=10.8,~E_F=0.1\mathrm{eV},~\alpha=10^{-4},~\mu_0H_e=0.3~\mathrm{T}$. Both the dissipation of plasmons and magnons are included as $\Gamma=0.1~\mathrm{meV},~\alpha=10^{-4}$.}\label{fig2}
\end{figure}

\textit{Reflection spectrum of the hybrid structures.}---
Now we continue to show how the hybrid excitation of surface plasmons and nutation spin waves can be detected by measuring the reflection spectrum of the hybrid DE$|$GRA$|$FM structure. Here we consider an Otto geometry and input a THz wave from medium 4 of the hybrid structure, where the permittivity of layer 4 is larger than that of layer  3, i.e. $\epsilon_4>\epsilon_3$, as shown in Fig. \ref{fig2}(a). Above a critical angle $\theta_c =\arcsin \sqrt{\epsilon_3/\epsilon_4}$, the input wave induces an evanescent wave in the medium 3, propagating toward the interface of graphene and ferromagnet. When the momentum and energy of the incident photons match that of the hybrid excitations, the nutation spin waves as well as the surface plasmons can be excited. The propagating of surface plasmons carries away electromagnetic energy and thus reduces the reflection rate of the incident wave, which can be measured. To be specific, here we input an electromagnetic wave $\mathbf{k}^{(i)}=(0,k_y,k_z)$ with $k_z=k_4 \cos \theta$, $k_y=-k_4 \sin \theta$, $\theta$ being the incident angle,  match the electromagnetic boundary conditions at the interfaces of media 4-3 and media 3-2, and then solve the reflection coefficient $R$ as \cite{YuanPRL2024}
\begin{equation}
R=\frac{k_2^2 (k_z \sinh(\kappa_3d_{DE})-i\kappa_3\cosh(\kappa_3d_{DE}))+\delta^- c_+}{k_2^2 (k_z \sinh(\kappa_3d_{DE})+i\kappa_3\cosh(\kappa_3d_{DE}))+\delta^- c_-},
\end{equation}
where $c_\pm \equiv (\mp\mu_0 \sigma \omega +k_z) \kappa_3 \cosh(\kappa_3d_{DE})\mp i(\kappa_3^2 \pm k_z \mu_0 \sigma \omega)\sinh(\kappa_3d_{DE})$ with $d_{DE}$ being the thickness of the dielectric medium 3.

Figure \ref{fig2}(b) shows the reflection spectrum of the hybrid system as a function of incident wave frequency. Clearly, a dip structure is observed at the nutation frequency (red line). We also plot the spin wave amplitude and find that the spin waves are maximally excited at the dip position (orange line). This coincidence shows that it is the nutation spin wave excitation that drags the motion of electrons in a 2D layer and enhances the electromagnetic loss of the hybrid system. As a comparison, the reflection rate of the hybrid DE$|$FM structure only has a small dip (black line) due to the absence of plasmon excitation, which highlights the significant role of plasmon-induced energy. Here the high conductivity of grpahene results in a strong plasmon excitation at magnetic resonance, making leading contribution to the reflection minimum, while the electromagnetic energy loss contributed by the magnetic resonance itself is secondary.
Figure \ref{fig2}(c) shows the one-to-one correspondence between the nutation time $\eta$ and dip position, which can be well described by our analytical result. Such feature implies that one can measure the position of the reflection minimum by THz spectroscopy and then quantitatively determines the nutation timescale.

\textit{Discussions and conclusions.}--Up till now, we focused on the nutation timescale around 0.1 ps or above, which may cover the nutation frequency of many known materials \cite{Mondal2023}. For materials with even smaller nutation time, the excitation surface plasmons will reach a long-wavelength infrared regime (tens of THz) to match the energy of nutation spin waves. Now the inter-band scattering of electrons will dominate the conducting behavior of graphene and the Drude model has to be modified as \cite{Falkovsky2007,Stauber2008}
\begin{equation}\label{conductivity_infrared}
\sigma= \frac{4\sigma_0E_F}{\pi(\Gamma - i\hbar \omega)} + \frac{i\sigma_0}{\pi}\ln \frac{2E_F-(\hbar \omega + i\Gamma)}{2E_F+(\hbar \omega + i\Gamma)},
\end{equation}
where the first and second terms come from the intra and inter-band scattering respectively. The dispersion equation of hybrid excitation in the absence of damping ($\alpha=0$, $\Gamma=0$) now becomes
\begin{equation}\label{dispersion_infrared}
\frac{k_2^2}{\delta^-} + \kappa_3 + \frac{\mu_0 \omega \sigma_0}{\pi} \left(\frac{4E_F}{\hbar \omega} + \ln \frac{2E_F-\hbar \omega}{2E_F+\hbar \omega}\right) =0. 
\end{equation}
This equation allows the existence of transverse electric (TE) plasmon mode in the frequency range of $1.67<\hbar \omega/E_F <2$ \cite{Mikhailov2007}, which is the condition that the third term of Eq. \eqref{dispersion_infrared} becomes negative. As an estimate, we take $\eta=0.01~\mathrm{ps}$, giving nutation frequency $\omega_+/(2\pi) \approx 15.9~\mathrm{THz}$, $E_F=0.037~\mathrm{eV}, 14.94~\mathrm{THz}<\omega_\mathrm{sp}/(2\pi) < 17.89~\mathrm{THz}$. Note that the graphene conductivity in this regime is vanishingly small due to the cancelation effect of inter and intra-band scattering in Eq. \eqref{conductivity_infrared}. Then the interacting channel of spin waves and surface plasmons mediated by the surface current $\mathbf{j}=\sigma \mathbf{E}_2$ is very weak and then the surface plasmon excitation does not play a significant role in the reflection spectrum. Nevertheless, one may still determine the strength of nutation utilizing the setup shown in Fig. \ref{fig2}(a) because the nutation spin wave emission will cause energy loss at the magnetic layer.

In summary, we have shown that THz nutation spin waves can hybridize with surface plasmons in a graphene$|$ferromagnet structure. Such a hybridization can take away electromagnetic energy and generate a measurable dip structure in the reflection spectrum of the hybrid system. The dip position gives a clear signature of spin inertial effects in a ferromagnet and can quantify the subsequent nutation timescale. Our proposal may help to identify the inertial dynamics in a wide class of magnetic materials and pave the way toward the study of the ultrafast dynamics of inertial spin wave-plasmon excitations. In principle, one can generalize our results to magnetic metals as well as antiferromagnetic systems. In magnetic metals,
spin oscillation can generate an eddy current at the metallic surface \cite{SilaevPRAp2022}, which will modify the magnetic damping as well as the electromagnetic boundary conditions. In an antiferromagnet, the normal spin wave frequency is also in the THz regime, comparable to the nutation frequency. The interplay of these two modes may generate interesting physics.


{\it Acknowledgments.}---We acknowledge Mikhail Cherkasskii for helpful discussions. This work is supported by the National Key R$\&$D Program of China (2022YFA1402700).

\section{Appendix: Derivation of the dispersion relation}
In this section, we show the mathematical details to derive the hybrid magnon-plasmon dispersion relation [Eq. \eqref{dispersion_hybrid} in the main text]. Considering the geometry of surface plasmon excitation in graphene-based hybrid structures shown in Fig. \ref{fig1}(a), the eigenmode of the hybrid system can be viewed as a surface electromagnetic wave \cite{YuanPRL2024} localized at the graphene layer. Here we consider the spectrum of a $s$-polarized or transverse electric (TE) surface electromagnetic wave which propagates along the in-plane $y-$axis and decays exponentially across the thickness direction ($z-$axis). In the dielectric medium 3, the general expression of the TE wave takes the form
\begin{equation}
\begin{aligned}
&\mathbf{E}_3=(E_{3,x},0,0)e^{ik_{3,y}y+\kappa_3z}, \\
&\mathbf{H}_3=(0,H_{3,y} , H_{3,z} )e^{ik_{3,y}y+\kappa_3z}.\\
\end{aligned}
\end{equation}
The magnetic field is connected to the electric field via the Maxwell equation $\nabla \times \mathbf{E}_3 =- \partial_t \mathbf{B}_3$ or $\mathbf{k}_3 \times \mathbf{E}_3 = \mu_0 \omega \mathbf{H}_3$.

Inside the magnetic medium 2, the finite thickness of the layer allows the coexistence of exponential decay and increase modes, i.e.
\begin{equation} \label{EM-ansartz}
\begin{aligned}
&\mathbf{E}_2=(E_{2,x}^{(-)},0,0)e^{ik_{2,y}y-\kappa_2z} + (E_{2,x}^{(+)},0,0)e^{ik_{2,y}y+\kappa_2z},\\
&\mathbf{H}_2=(0,H_{2,y}^{(-)}, H_{2,z}^{(-)}) e^{ik_{2,y}y-\kappa_2z} \\
&~~~~+ (0,H_{2,y}^{(+)}, H_{2,z}^{(+)}) e^{ik_{2,y}y+\kappa_2z}.
\end{aligned}
\end{equation}
Considering the magnetization dynamics governed by the LLG equation, which gives another relation between magnetization and magnetic field $\mathbf{M}_2=\chi \cdot \mathbf{H}_2$, the general Maxwell equation \eqref{MaxwellHM} now becomes
\begin{equation}\label{MaxwellHM2}
(\nabla^2 +k^2) \mathbf{H}_2- \nabla(\nabla \cdot \mathbf{H}_2) + k^2\chi \cdot \mathbf{H}_2 =0.
\end{equation}
Substituting the trial solution of magentic fields \eqref{EM-ansartz} into Eq. \eqref{MaxwellHM2}, we obtain
\begin{equation}
H_{2,y}^{(\pm)} = \frac{k_{2,y}k_{2,z} -i k_2^2 \nu }{k_{2,z}^2 - k_2^2 (1+\kappa)} H_{2,z}^{(\pm)}.
\end{equation}
Then the electric field is connected to the magnetic fields ($\nabla \times \mathbf{H}_2 = \epsilon_0 \epsilon_2 \partial_t \mathbf{E}_2$) as 
\begin{equation}
E_{2,x}^{(\pm)}= -\frac{1}{\epsilon_0 \epsilon_2 \omega} (k_{2,y}H_{2,z}^{(+)} - k_{2,z}H_{2,y}^{(+)})=-\frac{i\delta^{\pm}}{\epsilon_0 \epsilon_2 \omega}H_{2,y}^{(\pm)},
\end{equation}
where the coefficients $\delta^{\pm}$ are defined in the main text \eqref{deltapm}.

In the dielectric medium 1, the form of electromagnetic wave is similar to that of medium 3 as 
\begin{equation}
\begin{aligned}
\mathbf{E}_1&=(E_{1,x},0,0)e^{ik_{1,y}y-\kappa_1z}, \\
\mathbf{H}_1&=(0,H_{1,y} , H_{1,z} )e^{ik_{1,y}y-\kappa_1z}.\\
\end{aligned}
\end{equation}
The magnetic field is connected to the electric field via the relation $\mathbf{k}_1 \times \mathbf{E}_1 = \mu_0 \omega \mathbf{H}_1$.

The electromagnetic boundary conditions require the continuity of the tangential components of electric and magnetic fields at the two interfaces $z=0$ and $z=d$, respectively. At the interface $z=0$, the alternating electric current $j_x=\sigma E_{3,x}(z=0)$ generated by the plasmons in the graphene layer supplements the magnetic fields in the media 3 and 2. Overall, the four boundary conditions read
\begin{subequations}\label{boundary-equations}
\begin{align}
&E_{3,x}= E_{2,x}^{(-)} + E_{2,x}^{(+)},\\
&H_{3,y}-\sigma E_{3,x}= H_{2,y}^{(-)} + H_{2,y}^{(+)},\\
&E_{1,x}e^{-\kappa_1 d} = E_{2,x}^{(-)}e^{-\kappa_2 d} + E_{2,x}^{(+)}e^{+\kappa_2 d}, \\
&H_{1,y}e^{-\kappa_1 d} = H_{2,y}^{(-)}e^{-\kappa_2 d} + H_{2,y}^{(+)}e^{+\kappa_2 d}.
\end{align}
\end{subequations}
Here we have already imposed the requirement of in-plane momentum conservation, i.e. $k_{1,y}=k_{2,y}=k_{3,y}\equiv k_y$. By substituting the relation between electric and magnetic fields $\mathbf{k}_i \times \mathbf{E}_i = \omega \mathbf{B}_i$ into Eqs. \eqref{boundary-equations}, we derive a linear set of equations $\mathbf{A}\cdot \mathbf{X}=0$ with $\mathbf{X}=(H_{3,y},H_{2,y}^{(+)},H_{2,y}^{(-)},H_{1,y})$ and the coefficient matrix
\begin{equation}
\mathbf{A}=
\left ( \begin{array}{cccc}
  i \frac{k_2^2}{\kappa_3} & i \delta^{(-)}& i \delta^{(+)}&0\\
  1-i\frac{\mu_0 \omega \sigma}{\kappa_3} & -1  & -1 & 0 \\
  0 & i \delta^{(-)}e^{-\kappa_2 d} & i \delta^{(+)}e^{\kappa_2 d}&-i \frac{k_2^2}{\kappa_1} e^{-\kappa_1 d}\\
  0 & -e^{-\kappa_2 d} & -e^{\kappa_2 d}&e^{-\kappa_1 d}\\
\end{array} \right ).
\end{equation}
To guarantee a non-trivial distributions of electric and magnetic fields in the hybrid structure, it is required that $|\mathbf{A}|=0$, which readily recovers the dispersion relation \eqref{dispersion_hybrid} in the main text.

\begin{thebibliography}{}
\bibitem{Landau1935} L. Landau and E. Lifshitz, On the theory of the dispersion of
magnetic permeability in ferromagnetic bodies, in Perspectives in Theoretical Physics (Elsevier, New York, 1992) pp. 51–65.

\bibitem{Gilbert1956} T. L. Gilbert, A phenomenological theory of damping in 456
ferromagnetic materials, IEEE Trans. Magn. \textbf{40}, 3443 (2004).

\bibitem{Wegrowe2012} J. -E. Wegrowe and M.-C. Ciornei, Mangetization dynamics, gyromagnetic relation, and inertial effects. Am. J. Phys. \textbf{80}, 607 (2012).

\bibitem{KittelPR1948} C. Kittel, On the theory of ferromagnetic resonance absorption. Phys. Rev. \textbf{73}, 155 (1948).
\bibitem{Slon1996}J. C. Slonczewski, Current-driven excitation of magnetic multilayers. J. Magn. Magn. Mater. \textbf{159}, L1 (1996). 

\bibitem{Berger1996} L. Berger, Emission of spin waves by a magnetic multilayer traversed by a current. Phys. Rev. B \textbf{54}, 9353 (1996).

\bibitem{FukamiNN2016} S. Fukami, T. Anekawa, C. Zhang, and H. Ohno, A spin–orbit torque switching scheme with
collinear magnetic easy axis and current configuration. Nat. Nanotech. \textbf{11}, 621 (2016).

\bibitem{ChumakNP2015} A. V. Chumak, V. I. Vasyuchka, A. A. Serga, and B. Hillebrands, Magnon spintronics, Nat. Phys. \textbf{11}, 453 (2015).

\bibitem{YuanQM}  H. Y. Yuan, Y. Cao, A. Kamra, R. A. Duine, and P. Yan, Quantum magnonics: When magnon spintronics meets
quantum information science, Phys. Rep. \textbf{965}, 1 (2022).

\bibitem{YuPR2021} H. Yu, J. Xiao, H. Schultheiss, Magnetic texture based magnonics. Phys. Rep. \textbf{905}, 1 (2021).

\bibitem{Mougin2007} A. Mougin, M. Cormier, J. P. Adam, P. J. Metaxas, and J. Ferré, Domain wall mobility, stability and Walker breakdown in
magnetic nanowires. EPL. \textbf{78}, 57007 (2007).

\bibitem{Neeraj2021} K. Neeraj, N. Awari, S. Kovalev, D. Polley, N. Zhou, Hagström, S. S. P. K. Arekapudi, A. Semisalova, K. Lenz, B. Green, J.-C. Deinert et al., Inertial spin dynamics in ferromagnets, Nat. Phys. \textbf{17}, 245 (2021).
    
\bibitem{Unikand2022} V. Unikandanunni, R. Medapalli, M. Asa, E. Albisetti, D. Petti, R. Bertacco, E. E. Fullerton, and S. Bonetti, 
Inertialspin dynamics in epitaxial cobalt films, Phys. Rev. Lett. \textbf{129}, 237201 (2022).

\bibitem{YLi2015} Y. Li, A.-L. Barra, S. Auffret, U. Ebels, and W. E. Bailey, Inertial terms to magnetization dynamics in ferromagnetic thin films, 
Phys. Rev. B \textbf{92}, 140413(R) (2015).

\bibitem{Fähnle2011} M. Fähnle, D. Steiauf, and C. Illg, Generalized Gilbert equation including inertial damping: Derivation from an
extended breathing Fermi surface model, Phys. Rev. B \textbf{84}, 172403 (2011).

\bibitem{Bhat2012} S. Bhattacharjee, L. Nordström, and J. Fransson, Atomistic spin dynamic method with both damping and moment of
inertia effects included from first principles, Phys. Rev. Lett. \textbf{108}, 057204 (2012).

\bibitem{Danny2015} D. Thonig, J. Henk, and O. Eriksson, Gilbert-like damping caused by time retardation in atomistic magnetization dynamics. Phys. Rev. B \textbf{92}, 104403 (2015).

\bibitem{Kikuchi2015} T. Kikuchi and G. Tatara, Spin dynamics with inertia in metallic ferromagnets, Phys. Rev. B \textbf{92}, 184410 (2015).

\bibitem{Thonig2017} D. Thonig, O. Eriksson and M. Pereiro, Magnetic moment of inertia within the torque-torque correlation model. Sci. Rep. \textbf{7}, 931 (2017).
    
\bibitem{Mondal2017} R. Mondal, M. Berritta, A. K. Nandy, and P.M. Oppeneer, Relativistic theory of magnetic inertia in ultrafast spin dynamics, Phys. Rev. B \textbf{96}, 024425 (2017).

\bibitem{Mondal2018} R. Mondal, M. Berritta, and P.M. Oppeneer, Generalisation of Gilbert damping and magnetic inertia parameter as a series of higher-order relativistic terms. J. Phys.: Condens. Mater. \textbf{30}, 265801 (2018).

\bibitem{Mikhail2021} M. Cherkasskii, M. Farle, and A. Semisalova, Dispersion relation of nutation surface spin waves in ferromagnets. Phys. Rev. B \textbf{103}, 174435 (2021).

\bibitem{Mondal2023} R. Mondal, L. Rozsa, M. Farle, P. M. Oppeneer, U. Nowak, and M. Cherkasskii, Inertial effects in ultrafast spin dynamics. J. Magn. Magn. Mater. \textbf{579}, 170830 (2023).

\bibitem{HePRB2024} P.-B. He and M. Cherkasskii, Temporal and spatial attenuation of inertial spin waves driven by spin-transfer torques. Phys. Rev. B \textbf{110}, 174431 (2024).

    
\bibitem{Ciornei2011} M.-C. Ciornei, J. M. Rubí, and J.-E. Wegrowe, Magnetization dynamics in the inertial regime: Nutation predicted at short time scales.
Phys. Rev. B  \textbf{83}, 020410(R) (2011).

\bibitem{Titov2021} S. V. Titov, W. T. Coffey, Y. P. Kalmykov, M. Zarifakis, and
A. S. Titov, Inertial magnetization dynamics of ferromagnetic nanoparticles including thermal agitation, Phys. Rev. B
\textbf{103}, 144433 (2021).

\bibitem{Andres2022} J. Anders, C. R. Sait, and S. A. Horsley, Quantum Brownian
motion for magnets, New J. Phys. \textbf{24}, 033020 (2022).


\bibitem{Mario2024} M. G. Quarenta, M. Tharmalingam, T. Ludwig, H. Y. Yuan, L. Karwacki, R. C. Verstraten, and R. A. Duine, Bath-induced spin inerta. Phys. Rev. Lett. \textbf{133}, 136701 (2024).
    
\bibitem{YuanPRL2024} H. Y. Yuan and Yaroslav M. Blanter, Breaking surface plasmon excitation constraint via surface spin waves,  
Phys. Rev. Lett. \textbf{133}, 156703 (2024).

\bibitem{YuanPRB2025} H. Y. Yuan, Yaroslav M. Blanter, and H. Q. Lin, Strong and tunable coupling between antiferromagnetic magnons and surface plasmons, Phys. Rev. B \textbf{111}, 024422 (2025).

\bibitem{GMbook} A. G. Gurevich and G. A. Melkov, Magnetization oscillations and waves. CRC Press (1996).

\bibitem{Peresbook} P.A.D. Goncalves and N.M.R. Peres, An introduction to graphene plasmonics. World Scientific (2016).

\bibitem{Mikhailov2007} S. A. Mikhailov and K. Ziegler, New electromagnetic mode in graphene. Phys. Rev. Lett. \textbf{99}, 016803 (2007).

\bibitem{DE1960} J. R. Eshbach and R. W. Damon, Surface magnetostatic modes and surface spin waves. Phys. Rev. \textbf{118}, 1208 (1960).

\bibitem{DE1961} R. W. Damon and J. R. Eshbach, Magnetostatic modes of a ferromagnet slab. J. Phys. Chem. Solids. \textbf{19}, 308 (1961).

\bibitem{Stancil2021} D. D. Stancil and A. Prabhakar, Spin Waves: Problems and Solutions. 1st Edition, (Spinger Nature Switzerland AG, 2021).

\bibitem{Serga2010} A. A. Serga, A. V. Chumak, and B. Hillebrands, YIG magnonics. J. Phys. D: Appl. Phys. \textbf{43}, 264002 (2010).

\bibitem{WangPRL2020} H. Wang et al., Chiral spin-wave velocities induced by all-garnet interfacial Dzyaloshinskii-Moriya interaction in ultrathin yttrium iron garnet films. Phys. Rev. Lett. \textbf{124}, 027203 (2020).

\bibitem{MikhailPRB2020} M. Cherkasskii, M. Farle, and A. Semisalova, Nutation resonance in ferromagnets. 
Phys. Rev. B \textbf{102}, 184432 (2020).

\bibitem{MondalPRB2021} R. Mondal and A. Kamra, Spin pumping at terahertz nutation resonances. Phys. Rev. B \textbf{104}, 214426 (2021).
    
\bibitem{Kukh2015} S. M. Kukhtaruk and V. A. Kochelap, Semiclassical analysis of intraband collective excitations in a two-dimensional electron gas with Dirac spectrum, Phys. Rev. B \textbf{92}, 041409(R) (2015).
    
\bibitem{Silva2015} E. de Mello Silva, Dynamical class of a two-dimensional plasmonic Dirac system. Phys. Rev. E \textbf{92}, 042146 (2015).

\bibitem{Kukh2024} S. M. Kukhtaruk and V. A. Kochelap, Multivalued dispersion equation for coupling between plasmons and surface optical phonons in a graphene/polar-substrate system. Phys. Rev. B \textbf{110}, L241403 (2024).
    
\bibitem{Falkovsky2007} L. A. Falkovsky, and S. S. Pershoguba, Optical far-infrared properties of a graphene monolayer and multilayer. Phys. Rev. B \textbf{76}, 153410 (2007).
    
\bibitem{Stauber2008} T. Stauber, N. M. R. Peres, and A. K. Geim, Optical conductivity of graphene in the visible region of the spectrum. Phys. Rev. B \textbf{78}, 085432 (2008).

\bibitem{SilaevPRAp2022} M. Silaev, Anderson-HIggs mass of magnons in superconductor-ferromagnet-superconductor systems. Phys. Rev. Applied. \textbf{18}, L061004 (2022).

\end{thebibliography}
{}

\end{document}